\documentclass[preprint,12pt]{elsarticle}

\usepackage{graphicx}
\usepackage{amssymb}
\usepackage{lscape}
\usepackage{multirow}
\usepackage{color}
\usepackage{listings}
\usepackage{tabularray}
\usepackage{amsmath}
\usepackage{ragged2e}
\usepackage{svg}
\definecolor{dkgreen}{rgb}{0,0.6,0}
\definecolor{gray}{rgb}{0.5,0.5,0.5}
\definecolor{mauve}{rgb}{0.58,0,0.82}
\usepackage{geometry}
\usepackage{enumitem}
\usepackage{multirow}
\usepackage[]{color}
\usepackage{soul}
\newcommand{\hlr}[1]{{\colorbox{red}{#1}}}

\newcommand{\hlg}[1]{{\colorbox{green}{#1}}}
\newcounter{bla}

\journal{Computer Physics Communications}

\begin{document}

\begin{frontmatter}



\title{\texttt{CNUCTRAN}: A program for computing final nuclide concentrations using a direct simulation approach}


\author[a]{K A Bala}
\author[a]{M R Omar\corref{M R Omar}}
\author[a]{J Y H Soo\corref{J Y H Soo}}
\author[a]{W M H Wan Mokhtar}

\cortext[M R Omar] {First corresponding author.\\\textit{rabieomar@usm.my} }
\cortext[J Y H Soo] {Second corresponding author.\\\textit{johnsooyh@usm.my} }
\address[a]{School of Physics, Universiti Sains Malaysia, 11800 USM, Penang, Malaysia.}

\begin{abstract}
It is essential to precisely determine the evolving concentrations of radioactive nuclides within transmutation problems. It is also a crucial aspect of nuclear physics with widespread applications in nuclear waste management and energy production. This paper introduces \texttt{CNUCTRAN}, a novel computer program that employs a probabilistic approach to estimate nuclide concentrations in transmutation problems. \texttt{CNUCTRAN} directly simulates nuclei transformations arising from various nuclear reactions, diverging from the traditional deterministic methods that solve the Bateman equation using matrix exponential approximation. This approach effectively addresses numerical challenges associated with solving the Bateman equations, therefore, circumventing the need for matrix exponential approximations that risk producing nonphysical concentrations. Our sample calculations using \texttt{CNUCTRAN} shows that the concentration predictions of \texttt{CNUCTRAN} have a relative error of less than $0.001\%$ compared to the state-of-the-art method, CRAM, in different test cases. This makes \texttt{CNUCTRAN} a valuable alternative tool for transmutation analysis.
\end{abstract}

\begin{keyword}
Burnup Equation; Nuclide Concentrations; probabilistic approach; \texttt{CNUCTRAN} code

\end{keyword}

\end{frontmatter}



{\bf \noindent PROGRAM SUMMARY}

\begin{small}
\noindent
{\em Program Title: \texttt{CNUCTRAN}}                                          \\
{\em CPC Library link to program files:}  \\
{\em Developer's repository link:} https://github.com/rabieomar92/cnuctran/releases \\
{\em Licensing provisions: MIT software license}  \\
{\em Programming language: C++}                                   \\
{\em Nature of problem: \texttt{CNUCTRAN} simulates the transmutation of various nuclides such as decays, fissions, and neutron induced reactions using a direct simulation approach. It has the capability of predicting the final concentration of a large system of nuclides altogether after a specified time step, $t_f$.}\\
{\em Solution method: \texttt{CNUCTRAN} works based on the novel probabilistic method such that it does not compute the final nuclide concentrations by solving Bateman equations. Instead, it statistically tracks nuclide transformations into one another in a transmutation problem. The technique encapsulates various possible nuclide transformations into a sparse transfer matrix, \(\mathcal{T} \), whose elements are made up of various nuclear reaction probabilities. Next, \(\mathcal{T} \) serves as a matrix operator acting on the initial nuclide concentrations, \( \mathbf{y}(0) \), producing the final nuclide concentrations, \( \mathbf{y} \).
}\\
   \\
 \section{Introduction}
\noindent Understanding the changing concentrations of radioactive nuclides within a nuclear system is crucial for various applications in nuclear science. This knowledge underpins reactor analysis for optimal design and operation, the development of effective waste management strategies, the advancement of medical imaging techniques, and the assurance of robust radiation safety protocols. A set of linear differential equations governing the transformation of nuclides is central to understanding and predicting this phenomenon \cite{Omar2022}. These equations are commonly known as the Bateman equations, or alternatively burnup equations.

Determining the concentrations of nuclides with precision using the given equation has posed a significant challenge due to its complexity. Researchers have devised numerous codes and algorithms to tackle this issue. Most of these codes rely on a family of deterministic solvers known as the matrix exponential methods. For instance, ORIGEN \cite{Origen1980} was initially developed to address burnup-related problems and reactor physics applications. Some other codes that have burnup calculation capability include MONTEBURNS \cite{Monteburns}, SERPENT2 \cite{Serpent2015}, CINDER \cite{Cinder1995}, SCALE-4 \cite{Bowman1995}, and ALEPH2 \cite{Aleph2007}.

The fundamental principle underlying these codes is their ability to solve the system of Bateman equations. However, it's worth noting that these calculations tend to be computationally intensive and time-consuming, particularly when applied to extensive reactor analyses. A burnup problem becomes more challenging when more nuclides are included in the calculation, with half-lives ranging from $10^{-30}$s to $10^{30}$s. It makes the associated Bateman equations a stiff system of differential equations. Recently, IMPC-Burnup2.0 \cite{Zhao2021} was developed, employing the Transmutation Trajectory Analysis (TTA) \cite{Cetnar2021} and the Chebyshev Rational Approximation Method (CRAM) \cite{Pusa2013} to address the Bateman equations and perform analyses of accelerator-driven sub-critical systems (ADS). Other noteworthy calculation programs, such as ISOBURN \cite{Tavakkoli2021} and ONIX \cite{Lanversin2021}, utilize advanced algorithms to tackle burnup-related challenges. \hlr{Another notable improved explicit method is also} available, namely the Okamura Explicit Method (OEM). OEM modifies the matrix exponential method to achieve faster depletion calculations in dynamic nuclear fuel cycle simulations \cite{Okamura2022}.

In contrast, some of the previously mentioned codes employed conventional techniques that necessitate the utilization of ordinary differential equation (ODE) solution methods for solving the Bateman equations effectively. The Backward Differential Formulas (BDF) and the Runge-Kutta method are the commonly used ODE solution methods. For instance, the fifth-order implicit Runge-Kutta (IRK) method, integrated into the RADAU5 module within the ALEPH2 Monte Carlo burnup code, has demonstrated considerable success \cite{Stankovskiy2012}. Similarly, the BDF method, originally designed for solving first-order ordinary differential equations, has been incorporated through the LSODE package in the FISPACT-II code \cite{sublet2017}. Both the IRK and BDF methods find widespread application in addressing burnup problems. Nonetheless, it's worth noting that the IRK method requires careful adjustments due to potential limitations imposed by the size of the burnup matrix. On the other hand, the BDF method faces challenges associated with numerical solution stability and computational speed \cite{Stankovskiy2012}.

Our previous studies \cite{Omar2022, Omar2024} discovered the general statistical model that describes the random selection of transmutation events at the nucleus level. This was achieved by formulating a general probability distribution, facilitating the likelihood of occurrence for each enumerated nuclear reaction event. Additionally, an innovative probabilistic approach was devised to estimate solutions for burnup problems. \hlr{The proposed technique has a similar} approach to explicit methods as both of them rely on time discretization. However, the difference lies in the fact that the technique does not solve any differential equations. Unlike the conventional methods, the proposed approach leverages reaction probability calculations, therefore we coin the proposed method as the probabilistic approach. 

This paper presents a novel computer program named \texttt{CNUCTRAN}, which distinguishes itself by adopting a probabilistic approach in contrast to the conventional method relying on matrix exponential calculations. \texttt{CNUCTRAN}, implemented in C++, simulates nuclei transformations due to reactions such as decays, fissions, and neutron absorptions. The primary purpose of \texttt{CNUCTRAN} is to explore the feasibility of using the probabilistic approach to solving nuclide burnup problems. The probabilistic method may sound stochastic, but in reality, the technique does not incorporate any means of random sampling, and the final concentration produced is not contaminated with random errors. 
In contrast, other modern techniques such as TTA and CRAM employ different deterministic computational approaches. The key advantage of the probabilistic approach is that it avoids the risk of producing negative nuclides concentrations posed by CRAM. This is supported by the fact that CRAM has an inherent drawback in that it does not guarantee positivity of nuclide concentrations \cite{Muller2018, Hykes2013, Lahaye2017}. 
 
Furthermore, the probabilistic approach enables \texttt{CNUCTRAN} to perform a direct simulation of transmutation processes. In this sense, it meticulously monitor the transformations of nuclides within a burnup problem, maintaining the fidelity of the solution output. Also, our approach delves into the fundamental probabilistic principles underlying nuclide transmutation problems, tracing the temporal evolution of nuclide concentrations within a specified time interval, as elaborated in Section 2. Following the resolution of various nuclear transmutation challenges, our findings indicate that \texttt{CNUCTRAN} achieves final nuclide concentrations with a relative accuracy exceeding $10^{-5}$ when compared to benchmark problems \cite{lago2017}. This underscores \texttt{CNUCTRAN}'s readiness for burnup calculations across various analyses.

We organized the paper as follows. Section 2 covers the theoretical background as well as the theoretical method used in \texttt{CNUCTRAN}. Section 3 discusses the detailed description of the framework and method employed in the \texttt{CNUCTRAN} code system. Section 4 and 5 contains the verification and conclusion, respectively.
\section{Methodology} \label{section_method}
\subsection{The Bateman Equation} \label{sectbateman}
\noindent Consider a system with $I$ number of nuclides undergoing transmutations, thus transforming into one and another according to a specific transmutation chain. Let $i$ be the index of nuclide of interest and $\ell$ be the index representing other nuclides exclusive of $i$. The rate of change of the concentration of nuclide $i$ is given by the associated Bateman equation,
\begin{equation} \label{bateman}
	\frac{dy_i}{dt} = -\lambda_{i} y_i + \sum_{\ell=1}^I Y_{\ell\rightarrow i} \lambda_{\ell\rightarrow i} y_\ell
\end{equation}
where $y_i$ and $y_\ell$ are the concentration of nuclide-$i$ and nuclide-$\ell$, respectively; $\lambda_i$ is the total reaction rate constant of nuclide-$i$ causing its removal from the system; $\lambda_{\ell\rightarrow i}$ is the total reaction rate constant of nuclide-$\ell$ causing the transformation of nuclide-$\ell$ into nuclide-$i$; and, $Y_{\ell\rightarrow i}$ is the production yield of nuclide-$i$ due to the transformation of nuclide-$\ell$ into nuclide-$i$.

Eq. (\ref{bateman}) forms a system of $I$ linear differential equations dependent on one another. The solution of Eq. (\ref{bateman}) is the concentration of nuclide-$i$. For instance, one can define the concentration as the number of nuclei, or the percentage fraction. The LHS of the equation is balanced with the terms on the RHS. Here, each term describes the various transmutations of nuclides that lead to the production and removal of nuclide-$i$ within the system.
 
In particular, the first term on the RHS includes the natural decay and various  reactions that transform nuclide-$i$ into other nuclides, causing its disappearance from the system. In contrast, the second term of the RHS describes the formation of nuclides-$i$ via different transmutation reactions of all nuclides, including their natural decay reaction. Alternatively, if we define a nuclide concentration column matrix, $\mathbf{y} = \left(\begin{matrix}
	y_1 & y_2 & \cdots & y_I
\end{matrix}\right)^{\mathrm{T}}$, Eq. (\ref{bateman}) allows us to form the compact matrix form of a set of Bateman equations encompassing all $I$ nuclides,
\begin{equation}
	\frac{d\textbf{y}}{dt} = \mathcal{A}\textbf{y} 
\end{equation}
where $\mathcal{A}$ is the transmutation matrix such that its elements correspond to all coefficients defined on the RHS of Eq. (1), and of course, for all nuclides. Due to the nature of nuclear transmutations, $\mathcal{A}$ is a sparse $I\times I$ square matrix such that most of its off-diagonal elements are zeros. Also, $\mathbf{y}$ is a column vector corresponding to the concentration of all $I$ nuclides defined in the system. $\mathcal{A}$ is a matrix with an extraordinarily large norm, approximately on the order of $10^{21}$. This large magnitude of $\mathcal{A}$ implies significant numerical difficulties, as stiff systems exhibit varying timescales that make them particularly sensitive to numerical instability caused by short-lived nuclides. The solution to Eq. (1) demands specialized numerical methods capable of handling the stiffness, such as implicit techniques, to ensure accurate and reliable results.

\subsection{Conventional Method of Solution}
\noindent Obtaining the solution of Eq. (2) seems to be straightforward where the separation of variables is applicable, giving,
\begin{equation} \label{matrixexp}
	\textbf{y} = \exp(\mathcal{A}t) \textbf{y}_0 ,
\end{equation}
where $\textbf{y}_0$ is a list of initial concentration of nuclides considered in the calculation. Unfortunately, evaluating the matrix exponential in Eq. (\ref{matrixexp}) is difficult. Naively, one may approximate the matrix exponential by using a Taylor series,
\begin{equation} \label{matrixexp2}
	\exp({\mathcal{A}t}) \approx \sum_{k=0}^{\infty} \frac{1}{k!} \mathcal{A}^k t^k 
\end{equation}
However, burnup problems are stiff, such that $\mathcal{A}$ consists of elements with minimal and significant numeric values. Thus, evaluating Eq. (\ref{matrixexp2}) for a transmutation matrix is impractical and does not always yield an accurate and reliable result. Another potential method is the Pad\'e approximation of the matrix exponential, which unfortunately suffers the similiar drawback as the Taylor series approximation, but with a lesser degree.
 
Pusa \cite{pusa2016} discovered that the eigenvalues of $\mathcal{A}$ lie near the negative real axis, thus, enabling the approximation of $	\exp({\mathcal{A}t})$ via the Chebyshev approximants. Here, the technique is known as the Chebyshev Rational Approximation Method (CRAM), where the matrix exponential approximation is given by,
\begin{equation}\label{cram}
	\exp({\mathcal{A}t}) = a_0 + 2\mathrm{Re}\left[\sum_{i=1}^{k/2} (a_i{\mathcal{A}t - \theta_i \textbf{I}})\right],
\end{equation}
where $a_i$ and $\theta_i$ are the complex numbers published by Pusa \cite{pusa2016} for the approximation order $k \in\{16, 32, 48\}$. Here, the computation of Eq. (\ref{cram}) requires solving $k/2$ linear systems. Due to the sparsity pattern of the burnup matrix, the linear systems can be solved efficiently using sparse Gaussian elimination \cite{Pusa2013}. To date, CRAM is the most practical and can handle almost all burnup problems.

\subsection{The algorithm}
\begin{figure}
	\centering
	\includegraphics[width=0.5\linewidth]{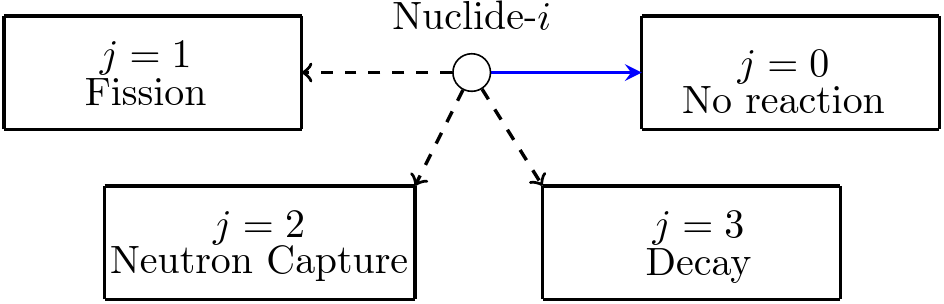}
	\caption{A nucleus of nuclide-$i$ can experience a total of $J$ probable reactions, however only one of these reaction will be randomly selected (blue arrow).}
	\label{fig:rxns}
\end{figure}

\noindent Suppose we define a transmutation event as an occurrence of a certain reaction that transform a parent nucleus into a daughter nucleus. The method begins with the establishment of the probability distribution of transmutation event occurring within a short time interval $\Delta t$. Alternatively we can ask about the probability that a nucleus to transform into another within $\Delta t$ via a certain reaction. The support of the probability distribution is the distinct type of possible reactions to be experienced by the nucleus. In this discussion, the same indexing convention as in Section \ref{sectbateman}  is used for $i$ and $\ell$, representing the parent and daughter nuclides, respectively. Now, it is convenient to introduce $j\in\{ 1, 2, \cdots, J\}$ as the reaction index identifying all distinct types of possible reactions to be experienced by nuclide $i$. We may also have to use the convention $j=0$, representing no reaction, and this is because it can also be considered as an event where a nucleus transforms into itself. Each nuclide can experience $J$ possible transmutation reactions that could transform itself into the daughter nuclide (see Fig. \ref{fig:rxns}). These reactions are probable, but in the end only one of them will be selected to occur. According to \cite{Omar2022}, the probability of a nucleus of nuclide $i$ undergoing reaction $j$ is given by,
\begin{equation} \label{first_dist_form}
	\pi_{i}(j; \Delta t) = \gamma_i \prod_{j'=1}^{J_i} \left\{\delta_{jj'} + (-1)^{\delta_{jj'}} e^{-\lambda_{j'} \Delta t}\right\}
\end{equation}
where $\gamma_i$ is the normalization constant, $\delta_{jj'}$ is the Kronecker delta and $\lambda_{j'}$ is a the reaction rate constant in s$^{-1}$, representing the intensity of the $j'$-th probable reaction to be experienced by nuclide $i$. The value of $\gamma_i$ is given by,
\begin{equation}
	\gamma_i = \left(\sum_{j=0}^{J_i} \tilde{\pi}_{i}(j; \Delta t)\right)^{-1}
\end{equation}
where $\tilde{\pi}_{i}(j; \Delta t)$ is the un-normalized form of the distribution function given in Eq. (\ref{first_dist_form}). With some simple algebraic manipulations, $\gamma_i$ can be evaluated and the complete form of the discrete probability distribution is given by:
\begin{equation} \label{discrete}
	\pi_i(j; \Delta t) =
	\begin{cases}
		\frac{\displaystyle e^{\lambda_{j} \Delta t} - 1}{\displaystyle 1+ \sum_{j'=1}^{J} \left (e^{\lambda_{j'} \Delta t } -1 \right)}  & \text{for } j \in \left \lbrace 1, 2, \dots, J \right \rbrace  \\
		\frac{\displaystyle 1}{\displaystyle 1+\sum_{j'=1}^{J}  \left (e^{\lambda_{j'} \Delta t } -1 \right)} & \text{for } j = 0 \text{ (no reaction)}
	\end{cases}
\end{equation}
Let $t$ be the current time and $t+\Delta t$ be the future time, thus forming an instantaneous time interval  $(t,t+\Delta t)$. Generally, the concentration of nuclides $i$ during the future time, $t+ \Delta t$, is contributed by two factors. Firstly, it is contributed by the un-reacted nuclides $i$ within the interval. Secondly, it is contributed by the transmutation of other nuclides that occur within the interval, which leads to the production of nuclide $i$. Therefore, the concentration of nuclide $i$ after the next sub-step, $t+\Delta t$ , can be expressed as,
\begin{equation} \label{p1}
	{y}_i(t+\Delta t) = \sum_{\ell=1}^{I} \tau_{\ell\to i} y_\ell(t) 
\end{equation}
where here we define $\tau_{i\to i}$ as a scaling factor representing nuclide $i$ not undergoing a transmutation event, and $\tau_{\ell\to i}$ as a scaling factor representing any nuclide $\ell$ transforming into nuclide $i$ via various types of transmutation reactions. Briefly, $\tau$ is a factor that adjusts and updates the instantaneous value of nuclide concentrations during $t$ into $t + \Delta t$. By definition,
\begin{equation} \label{p2}
	\tau_{\ell\to i} = \sum_{\forall j \in \mathcal{R}_\ell} Y_{j,\ell \rightarrow i}\pi_\ell(j;\Delta t) 
\end{equation}
where $\mathcal{R}_\ell$ is a set of transmutation events (the $j$'s) that transform nuclide $\ell$ into daughter nuclide $i$, and $Y_{j,\ell\rightarrow i}$ is the production yield of nuclide $i$ due to the reaction $j$. If removal event $j$ is a fission reaction, then $\pi_\ell$ in Eq.(\ref{p2}) must be scaled to the fission yield of the daughter nuclide $i$. Notice also that $Y_{0,i\rightarrow i}=1$, because reaction $j=0$ is defined as no reaction. Eq.(\ref{p1}) can be systematically written in matrix form via the definition given in Eq.~(\ref{p2}). Here, by considering all $I$ nuclides in a burnup problem,
\begin{equation}
	\left(\begin{array}{c}
		y_1(t+ \Delta t)  \\
		y_2(t+ \Delta t) \\
		\vdots\\
		y_I(t+ \Delta t)  \\
	\end{array}\right)=\left(\begin{array}{cccc}
	\tau_{1 \rightarrow 1} & \tau_{2\rightarrow 1} & \cdots & \tau_{I\rightarrow 1} \\
	\tau_{1\rightarrow 2} & \tau_{2\rightarrow2} & \cdots & \tau_{I\rightarrow 2} \\
	\vdots & \vdots & \ddots & \vdots \\
	\tau_{1\rightarrow I} & \tau_{2\rightarrow I} & \cdots & \tau_{I\rightarrow I} \\
	\end{array}\right)\left(\begin{array}{c}
	y_1(t)  \\
	y_2(t) \\
	\vdots\\
	y_I(t)  \\
	\end{array}\right)
\end{equation}
Or simply,
	
	\begin{equation}
		\textbf{y}(t + \Delta t) = \mathcal{T}(\Delta t) \textbf{y}(t) \
\end{equation}

We note that $\mathcal{T}$ is constructed based on the sub-step interval, $\Delta t$. Eq. (11) is a recursion relation where the future nuclide concentrations, $\textbf{y}(t+\Delta t)$, can be obtained by transforming the nuclide concentrations of the previous time interval via the transfer matrix, $\mathcal{T}$. Here,  $\mathcal{T}$ serves as the transformation operator. The column indices of $\mathcal{T}$ correspond to the various parent nuclides, and the row indices of $\mathcal{T}$ correspond to the different daughter nuclides. Notice that the diagonal elements of square matrix $\mathcal{T}$ are probabilities of no transmutation events occurring.

Suppose that the evolution of nuclide concentration is monitored from the initial time, $t=0$ s, to the final time, $t=t_f$. Then, a small-time interval, $\Delta t$, is chosen such that the total number of sub-steps is given by $\nu=t_f/\Delta t$. Therefore, $y(t)$ can be obtained by repeating Eq. (11), for $\nu$ times, giving,

\begin{equation}
	\textbf{y}(t_f) = \textbf{y}(\nu\Delta t) = \mathcal{T}^\nu \textbf{y}(0) 
\end{equation}
where $\textbf{y}(0)$ is the initial nuclides concentration at $t=0$ s. Here, the smaller the value of $\Delta t$, the larger the number of sub-steps, thus the better the accuracy. However, such a computation with small $\Delta t$ requires arbitrary precision arithmetic to avoid the floating-point error. Evaluating $ \mathcal{T}^\nu$ requires $\nu$ repeated multiplications of $\mathcal{T}$. Such a naïve approach will incur an unreasonably long CPU time, especially when $\nu$ is large. Therefore, to improve the method's accuracy and avoid unnecessary zero matrix multiplication, we included two approaches: exponentiation by squaring helps to evaluate $\mathcal{T}^\nu$ with only $\log_2 \nu$ multiplications \cite{Gordon1998}. Secondly, a sparse matrix multiplication (SpMM) algorithm \cite{Gustavson1978} was implemented since it involves matrix multiplication.
We note that $ \mathcal{T}^\nu$ can be efficiently calculated by repeated squaring of $ \mathcal{T}$.
\subsection{Approximation Order ($n$)} \label{calculation_order}
\noindent Since the probabilistic method requires $\nu$ self-multiplication of the transfer matrix,  $\mathcal{T}$, we implemented the repeated squaring technique to speed up the computational time. Sometimes, this technique is coined as \textit{exponentiation by squaring}. Now, we let the number of sub-steps to be equivalent to $\nu$ and has the following form,
\begin{equation}
	\nu = 2^k \
\end{equation}
where $k$ is the total number of repeated self-multiplications of $\mathcal{T}$. Our objective is to obtain the suitable value of $k$, such that the sub-step interval $\Delta t$ is in order of $10^{-n}$. Here, $n$ is known as the approximation order, and the final time step, $t_f$, of the calculation is given below:
\begin{equation}
	t_f \approx 2^k \Delta t = 2^k \times 10^{-n} \
\end{equation}
Solving for $k$,
\begin{equation}\label{flooroper}
	k = \lfloor \log_2 (10^n t_f) \rfloor \
\end{equation}
The floor operation in Eq. (\ref{flooroper}) is needed because $k$ must be a positive integer number. Also, $\Delta t$ must be corrected to compensate for the error due to the floor operation:
\begin{equation}
	\Delta t = \frac{t_f}{2^{\lfloor \log_2 (10^n t_f) \rfloor}} \
\end{equation}
Fig. \ref{flowchart} summarizes how the probabilistic method is implemented in \texttt{CNUCTRAN}. In short, \texttt{CNUCTRAN} accepts \texttt{input.xml} and \texttt{chain\_endfb71.xml} as its inputs. Then it constructs the transfer matrix, $\mathcal{T}$, and lastly it computes and stores the final concentration of nuclides in \texttt{output.xml}.

\begin{figure}
	\centering
	\includegraphics[width=1.0\textwidth]{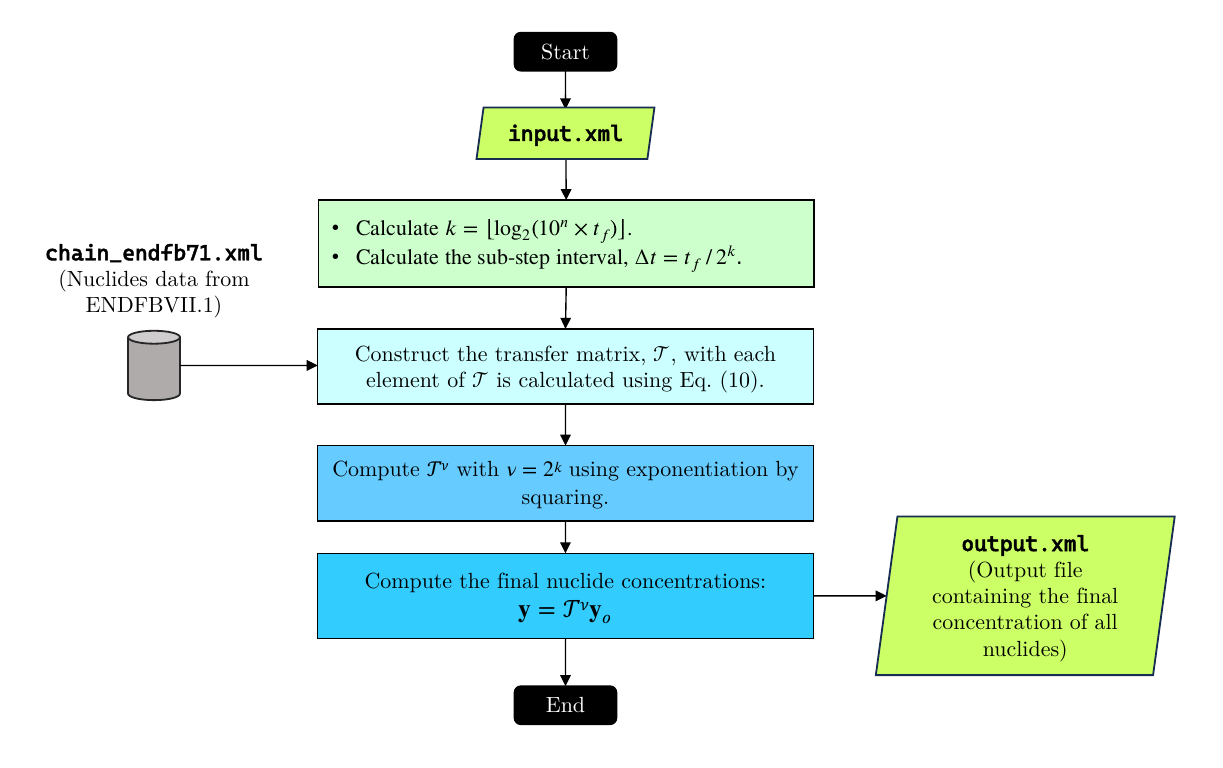}
	\caption{A flowchart representing how \texttt{CNUCTRAN} works.}
	 \label{flowchart}
\end{figure}

\section{Features of \texttt{CNUCTRAN}}
\subsection{\texttt{CNUCTRAN} Storage}
\noindent The code works by analyzing the reactions of all nuclides and compute their final concentrations after a given final time step, $t_f$. For proper functionality, \texttt{CNUCTRAN} requires two specific input XML files and two dynamic-linked libraries (\texttt{mpir.dll} and \texttt{mpfr.dll}) that facilitate arbitrary precision arithmetic. They are packaged together in the program folder and must coexist with the \texttt{CNUCTRAN} executable program within the same working directory to ensure error-free execution. Even so, the users are only required to prescribe two XML files. The first file is \texttt{input.xml}, an XML file that defines problem zones and computation settings. If the file has a different name, the program command line argument must specify the new name. The second XML file serves as the nuclides library, containing transmutation chains for various nuclides. It must include crucial details such as half-life values and reaction  products for various reactions such as $\alpha$, $\beta$, and $\gamma$ decays, as well as neutron-induced reactions such as $(n, \alpha)$, $(n, \gamma)$, and $(n,2n)$. By default, \texttt{CNUCTRAN} comes with a preset library named \texttt{chain\_endfb71.xml}. This library, derived from the Computational Reactor Physics Group (CRPG) at MIT, is based on the Evaluated Nuclear Data File (ENDF) version B-VII.1 \cite{Chadwick2011}.

\texttt{CNUCTRAN} utilizes XML files for structuring input data because XML format is well-understood by most experts. In contrast to many other nuclear-related codes that rely on arbitrary-format ASCII files "cards" for specifying the geometry, materials, and run settings, \texttt{CNUCTRAN} opts for a more organized approach.

\subsection{Input Specification} \label{input_section}
\noindent Before executing the code, a model definition representing the specific problem of interest is required. By default, \texttt{input.xml} contains the definition of the burnup problem that allows \texttt{CNUCTRAN} to construct the associated transmutation chain. The model requires the definition of one or more zones within a nuclear reactor or any other physical system containing the nuclides under investigation. The user's task is to articulate this model so that the code can produce results that resemble the actual physical burnup problem. A basic model comprises several essential components (refer to Figure \ref{input}): firstly, the description of zones, dividing the problem into spatial regions with distinct nuclide compositions. Secondly, various simulation settings instruct the code on precision digits, final time steps, calculation order, and other preferred options. The final component is the data output file, defined within the model.

The code employs XML markups written in a single input file with an \texttt{.xml} extension to capture these definitions. Further details describing the input XML tags that are used in the file are elaborated below and in Tables \ref{table_zone} and \ref{table_settings}. The structure of the input XML file must always begin with the \texttt{problem} parent tag. The child value of \texttt{<problem>} must at least contain one zone definition. Additionally, the \texttt{<zone>} tag should only have one attribute: the zone's name. All tags within the \texttt{<zone>} tag are considered child elements of \texttt{<zone>}. The specifications for all child tags within the \texttt{<zone>} tag are provided in Table \ref{table_zone}. In addition, Table \ref{table_settings} lists the specifications for all child tags within the \texttt{<simulation\_params>} tag, along with examples of how the setting of a given problem will be specified in the input file.

\begin{table}[h] 
	\caption{Specifications of all child tags within the \texttt{<zone>} tag.}
	\centering
	\small
	\begin{tblr}{
			width=\textwidth,
			hlines,
			vlines,
			colspec={X[3,l] X[1,l] X[3,l]},
			rowsep=1ex,
		}
		\textbf{Tag Name \& Information  }             & \textbf{Attributes} & \textbf{Description}                                                                                                                                                                                         \\
		\SetCell[r=3]{l}\texttt{species} \newline \newline
		\justifying 
		
		\noindent \textit{Purpose:} Lists all nuclide species in the calculation.
		
		\noindent\textit{Child Value:} The child value should be a list of nuclide names from the nuclides library, separated by newline, whitespaces, or commas. If \texttt{amin} and \texttt{amax} are specified, the program will disregard the given list of nuclide names. & \texttt{source}     & \justifying Sets the nuclides library location on the disk. The value should be the
		location of the \texttt{chain\_endfb71.xml} file.                                                                                                                             \\
		& \texttt{amin}       & \justifying Sets the minimum atomic number, $A_{\text{min}}>0$, to be included in the calculation.                                                                                                                               \\
		& \texttt{amax}       & \justifying Sets the maximum atomic number, $A_{\text{max}}>0$, to be included in the calculation.                                                                                                                               \\
		\SetCell[r=2]{l} \texttt{initial\_concentrations} \newline \newline \textit{Purpose}:
		Specifies the initial concentration of all nuclides involved in the calculation. 
		
		\noindent \textit{Child Value}:
		See \texttt{<concentration>} tag.
		 & \texttt{source}     & \justifying Sets the initial concentration XML file location on the disk. This file is the output XML file produced by any previous calculation.                                                                                                                         \\
		& \texttt{override}   & This attribute is helpful if the source attribute is prescribed. There are two allowed values:                                                                                                      \\
		\SetCell[r=2]{l} \texttt{concentration} \newline \newline \textit{Purpose}:
		Defines a nuclide concentration value.
		
		\textit{Child Value}:
		Not applicable.
		      & \texttt{species}    & Optional if source attribute of \texttt{initial\_concentrations} is not set.                                                                                                                                 \\
		& \texttt{value}      & The value of the concentration. The concentration unit can be anything, including mol, kg, g, unitless (nuclei count), etc. However, the unit must                                                  \\
		\texttt{reaction\_rates} \newline \newline \textit{Purpose}:
		Lists all neutron-induced reactions
		associated with various nuclides 
		involved in the calculation (optional).
		
		\textit{Child Value}:
		Empty or multiple \texttt{<reaction>} tag(s).
		        & n/a        & n/a                                                                                                                                                                                                 \\
		\SetCell[r=3]{l} \texttt{reaction}                & \texttt{species}    & The name of the target nuclide species undergoing the reaction.                                                                                                                                     \\
		& \texttt{type}       & The reaction type. The value can be fission, (n, gamma), (n, 2n), (n,a) depending on the prescribed nuclides data library.                                                                          \\
		& \texttt{rate}       & Reaction rate in s$^{-1}$. 
	\end{tblr}
	\label{table_zone}
\end{table}

\begin{table}[htbp]
	\centering
	
	\small
	\caption{Specifications of all child tags within the \texttt{<simulation\_params>}}
	\resizebox{\textwidth}{!}
	{
\begin{tabular}{|l|p{12cm}|}
	\hline
	{Tag Name} & {Child Value} \\
	\hline
	\texttt{<n>} & Sets the approximation order of the calculation. The precision is measured in terms of \(10^{-n}\). For more information on the approximation order, refer to Section \ref{calculation_order}. The child value must be an integer greater than 0. A value greater than ten is recommended. However, a higher value will increase the computational cost. \\
	\hline
	\texttt{<time\_step>} & Sets the final time step of the calculation in seconds. The child value must be a float number greater than 0. \\
	\hline
	\texttt{<precision\_digits>} & Sets the minimum number of accurate digits maintained in the arithmetic. The child value must be an integer greater than 30. A higher value will increase the computational cost. Depending on the problems, it is recommended to set the precision to \(>40\) for the approximation order, \(n>10\). The default value is 45. \\
	\hline
	\texttt{<output\_digits>} & Sets the number of decimal points to be displayed in the output. The child value must be an integer greater than 0. The default value is 16. \\
	\hline
	\texttt{<output>} & Sets the location of the output XML file on the disk. The child value must be a valid target file name of the XML output file. The default is \texttt{output.xml}. \\
	\hline
	\texttt{<verbosity>} & Sets the verbosity level. A higher verbosity means more console output messages will be displayed. A value of one enables a higher verbosity level. The default verbosity level is 0. \\
	\hline
\end{tabular}
}
\label{table_settings}
\end{table}

To better understand the input specifications, let us consider a burnup zone consists of nuclides with an atomic number between \(A = 100\) and \(A = 260\). An example \texttt{input.xml} for this example is shown in Fig. \ref{input}. The problem is a pure decay problem, and we are interested in calculating the final nuclide concentrations after 1 million years (\(\approx 3.1556926 \times 10^{13}\) seconds). The location of the nuclide data library is in the same working directory with the name \texttt{chain\_endfb71.xml}. The calculation order is \(n = 15\), and the number of precision digits is 50. The final nuclide concentrations are stored in \texttt{output.xml} within the same working directory. The initial concentration of Np-237 is \(1.0\) mol.

\begin{figure}[h] 
	\caption{An example \texttt{input.xml} for a single burnup zone, 1mol Np-237 undergoing pure decay. Tracking daughter nuclides with $100 \le A \le 260$.}
	\begin{lstlisting}[language=XML, label={lst:input1}]
		<problem>
		
			<!-- Zone definition.-->
			<zone name="myzone">
				<species source="chain_endfb71.xml" amin="100" amax="260" />
				<initial_concentrations>	
					<concentration species="Np237" value="1.0" />	
				</initial_concentrations>
				<reaction_rates>
					<!--This area is intentionally left blank as only decay reactions are considered. -->
				</reaction_rates>
			</zone>
			
			<!-- Simulation parameters.-->
			<simulation_params>
				<n>15</n>
				<time_step>31537110285757.0</time_step>
				<precision_digits>50</precision_digits>
				<output_digits>20</output_digits>
				<verbosity>1</verbosity>
				<output>.\output.xml</output>
			</simulation_params>
		
		</problem>
		
	\end{lstlisting}
	\label{input}
\end{figure}
\subsection{\texttt{CNUCTRAN} Output}
\noindent To execute the computation in the algorithm, users must have all input files prepared and run \texttt{\texttt{CNUCTRAN}.exe}. The executable requires two dynamic libraries, i.e., \texttt{mpfr.dll} and \texttt{mpir.dll}, to be placed in the same directory. Upon completion of the calculation, \texttt{CNUCTRAN} will generate an output XML file that contains the final nuclide concentrations. Users can manually specify the name and location of the output XML file through the input file. This can be done using the \texttt{<output>} tag (refer to Table \ref{table_zone}).

Additionally, users can adjust the verbosity level of the console output using the \texttt{<verbosity>} tag. The tag’s value can be set to 0, 1, or 2. A higher value indicates a higher verbosity level. If the verbosity level is set to zero, \texttt{CNUCTRAN} will suppress all progress messages. However, it will still display error messages. Setting the verbosity level to 1 allows \texttt{CNUCTRAN} to print essential messages to the console output, such as the final time step, the total number of nuclides involved in the calculation, and the CPU time required for the computation. 

An example output of the problem discussed in Section \ref{input_section} is given in Fig. \ref{output_demo}. Note that only the non-zero nuclide concentrations are presented.
\begin{figure} [h]
	\caption{Example output from \texttt{CNUCTRAN} in \texttt{output.out} file. There are a total of 2496 nuclide concentration entries, \hlg{however, only a few entries are displayed.}}
	\begin{lstlisting}
		CNUCTRAN v1.1 OUTPUT.
		final time step = 3.153711e+13s
		order (n) = 15
		total nuclides = 2496
		substep = 1.214765e-20s (111 sparse mults.)
		precision = 50 digits.
		Species Non-zero Concentration
		Tl205    yes       3.38415904648820962141e-15
		Tl209    yes       3.30493926518643221077e-14
		Pb209    yes       1.33285785761905294429e-10
		Bi209    yes       2.16421737645377106870e-01
		Bi213    yes       3.11306159296745811710e-11
		Po213    yes       4.67504829401650235053e-20
		At217    yes       3.67620562165477868849e-16
		Rn217    yes       4.30218490707587109346e-22
		Fr221    yes       3.34614381661456514015e-12
		Ra225    yes       1.46520124548860239498e-08
		Ac225    yes       9.83356550185507195873e-09
		Th229    yes       2.63631738725039137825e-03
		Pa233    yes       2.49362888205028377063e-08
		U233     yes       5.70305339777048435029e-02
		Np237    yes       7.23911361400001534277e-01
		:        :         :
		:        :         :
	\end{lstlisting}
	\label{output_demo}
\end{figure}
\section{Verification}
\noindent To verify the consistency of \texttt{CNUCTRAN}, we examined two  samples calculations for comparison. For all calculations using \texttt{CNUCTRAN}, the number of precision digits was set to 50 and the calculation order was set to $n=20$. The first sample calculation focused on the simple decay reaction resulting \hlg{from the neutron capture on Bi-209, and the transmutation chain involving the} important nuclides is given as follows:

\begin{equation*}
	{ }^{209} \mathrm{Bi} \xrightarrow{(n, \gamma)}{ }^{210} \mathrm{Bi} \xrightarrow{5.013 d}{ }^{210} \mathrm{Po} \xrightarrow{138.376 d}{ }^{206} \mathrm{Pb}
\end{equation*}

The initial concentration of Bi-209 is \(0.0165368\) mol. \hlg{Bi-209 is under a constant} neutron irradiation over the entire time step, experiencing \((n, \gamma)\) reaction with a rate constant of \(2.09 \times 10^{-11}\,\textrm{s}^{-1}\). \hlg{Theoretically, the reaction rate constant} for neutron-induced reactions is given by

\begin{equation}
	\lambda = \sigma \phi
\end{equation}
where $\phi$ is the average neutron flux (in cm$^{-2}$ s$^{-1}$) and $\sigma$ is the microscopic neutron cross section (in cm$^{2}$). For nuclear reactor simulations, $\phi$ can be readily obtained from any neutron transport codes. \texttt{CNUCTRAN} requires the user to specify the reaction rate constant of all neutron-induced reactions. For decay reactions, $\lambda$ is automatically calculated by \texttt{CNUCTRAN} using the following well known formula,

\begin{equation}
	\lambda_{\text{decay}} = \frac{\ln 2}{t_{1/2}}
\end{equation}
where $t_{1/2}$ is the half-life value obtained from the nuclides data (\texttt{chain\_endfb71.xml}).

We monitored the progression of the problem in four distinct final time steps ($t_f$): 30 days, one year, 30 years, and 1 million years. Table \ref{tbl:spec_sample1} provides comprehensive data concerning the half-lives, and decay constants associated with the important nuclides within the linear chain. We evaluated the concentrations of these nuclides, denoted as \(y(t)\), employing the probabilistic approach discussed in Section \ref{section_method}. We also define the important nuclide, which has a final nuclide atomic fraction of $\ge10^{-15}$. The results are presented in Fig. \ref{fig:sample_1}, depicting the plot of relative errors between \texttt{CNUCTRAN} and CRAM versus nuclide concentrations for the first sample calculation. Specifically, we compare these results with CRAM of order 48 (\texttt{CRAM48}), and our findings show a good agreement between the probabilistic method and CRAM, with only one nuclide having a relative error $<10^{-5}$. \hlg{Notice that both \texttt{CNUCTRAN} and \texttt{CRAM48} calculation also determines the} concentration of unimportant nuclides, which typically has a value of less than $10^{-15}$.

The second sample calculation focuses on the combined fission, neutron captures and decay scenarios, specifically involving U-238 and U-235 nuclides within the atomic number range of \(A=150\) to \(A=264\). \hlg{The second sample calculation encompasses 2494 daughter nuclides} of U-235 and U-238. Therefore, including a full chain is not feasible due to the complex transformations between nuclides. \hlg{U-235 and U-238 are under constant neutron irradiation over the entire time step}. They \hlg{experience \((n, \gamma)\) reaction at a rate of \(1 \times 10^{-4}\,\textrm{s}^{-1}\)} and U-235 undergoes fission at a rate of \(1 \times 10^{-5}\,\textrm{s}^{-1}\). The parent nuclides, U-238 and U-235, begin with the initial atomic fractions of \(0.9925\) and \(0.0075\), respectively. To track the burnup process, we monitor the sample over the following final time steps ($t_f$): 30 days, one year, 30 years, and 1 million \hlg{years. Fig. \ref{fig:sample_2} visually} represents the relative errors in nuclide concentrations when utilizing \texttt{CNUCTRAN} for the second sample calculation. Similar to the first sample calculation, we compared the calculated solution obtained from \texttt{CRAM48}. 
\begin{table}[h]
	\centering
	\small
	\caption{Calculation parameters of Bi-209 Burnup problem (first sample calculation )}

	\begin{tabular}{@{}clccc@{}}
		\hline
		{$i$} & {Nuclide} & {$y_0$ (at/b-cm)} & {Half-life (s)} & {Decay Constant (\(\textrm{s}^{-1}\))} \\
		\hline
		1 & Bi-209 & 0.0165368 & \(6.00 \times 10^{26}\) & \(2.09 \times 10^{-11}\) \\
		2 & Bi-210 & 0.0 & \(4.33 \times 10^{5}\) & \(1.60 \times 10^{-06}\) \\
		3 & Po-210 & 0.0 & \(1.20 \times 10^{7}\) & \(5.80 \times 10^{-08}\) \\
		4 & Pb-206 & 0.0 & 0.00 & 0.00 \\
		\hline
	\end{tabular}

      \label{tbl:spec_sample1}
\end{table}
\begin{figure}[h]
	\centering
	\includegraphics[width=1.0\textwidth]{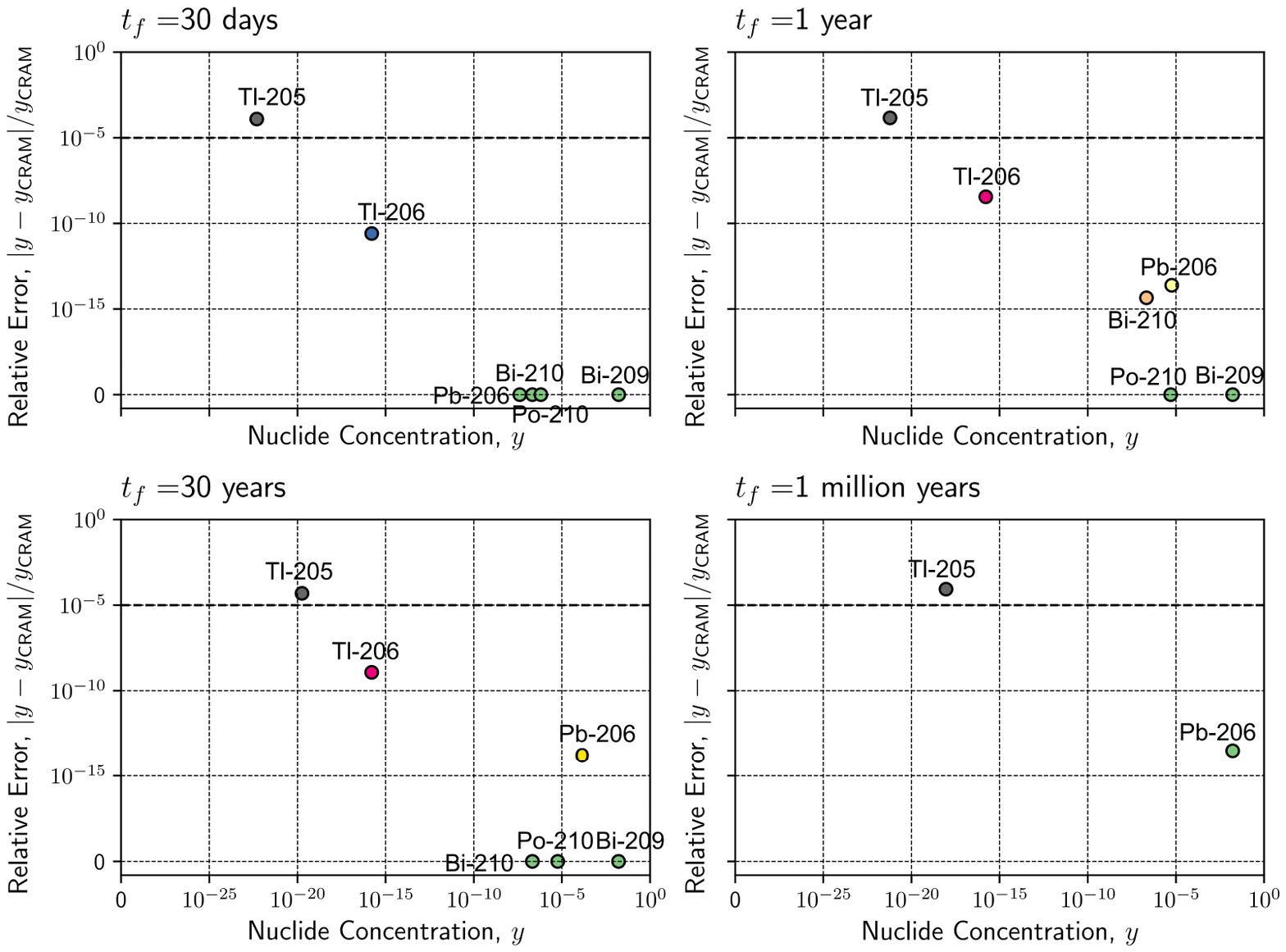}
	\caption{Relative errors versus the final nuclide concentrations of Bi-209 and its daughter nuclides after 1 month, 1 year, 30 years and 1 million years.}
	\label{fig:sample_1}
\end{figure}
\begin{figure}[h]
	\centering
	\includegraphics[width=1.0\textwidth]{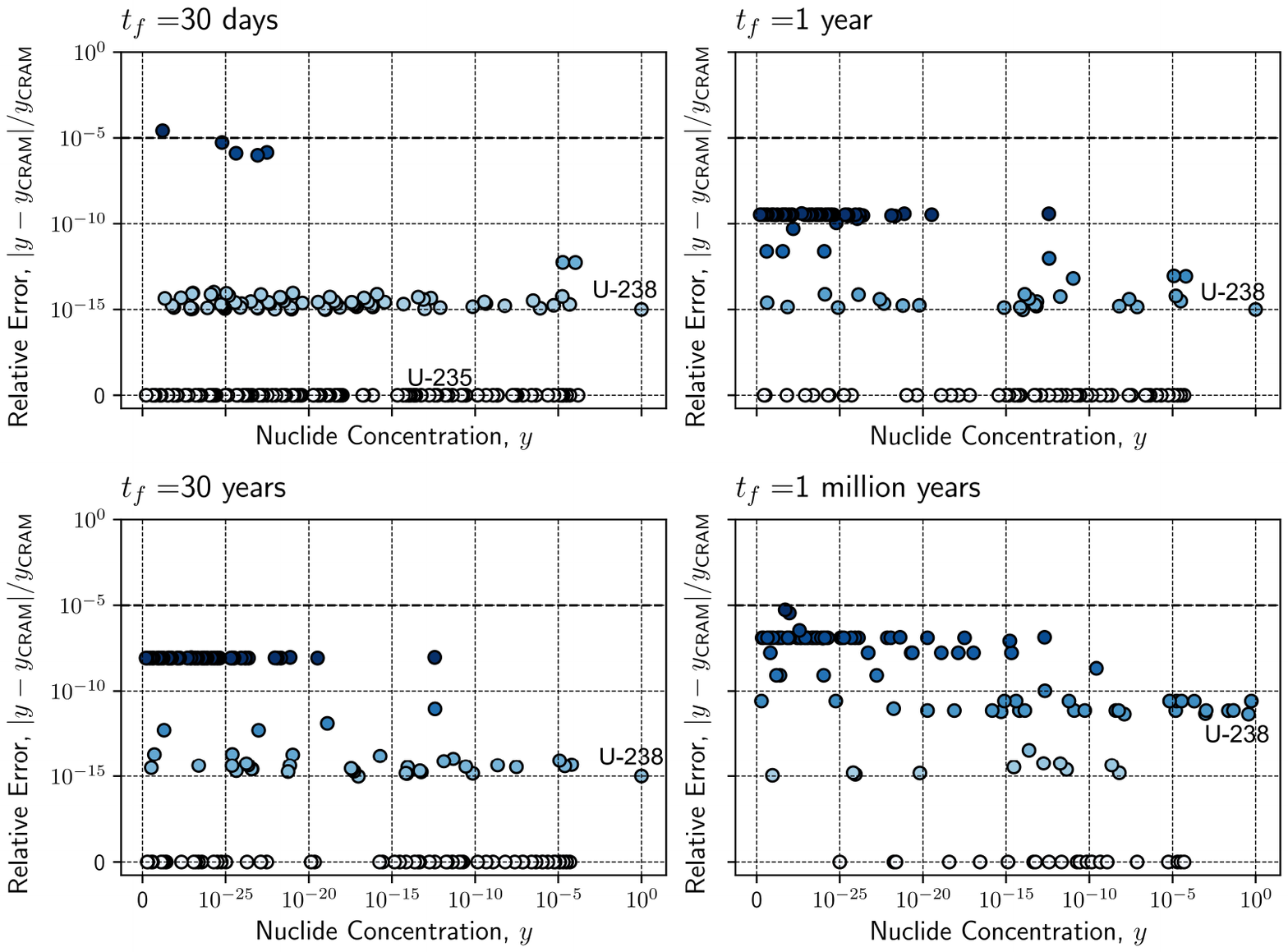}
	\caption{Relative errors versus the final nuclide concentrations of U-238, U-235 and their daughter nuclides after 1 month, 1 year, 30 years and 1 million years.}
	\label{fig:sample_2}
		
\end{figure}

Our analysis used the same nuclear data for all sample calculations. To assess the accuracy of our code, we employed different approaches. We compared the results with CRAM of order 48 (\texttt{CRAM48}) for the first and the second sample calculations. Interestingly, we observed a higher level of consistency with the probabilistic method where most final nuclides concentrations having a relative error of $ <10^{-5} (0.001\%)$. \texttt{CRAM48} was chosen because it is by far the most accurate and efficient state-of-the-art depletion calculation method. Fig. \ref{fig:sample_2}  visually represent the results for the second sample calculation.

Notably, the calculated solutions exhibit no significant differences from the reference solutions. The largest error observed is less than 0.01\%. The saturation of errors at $10^{-15}$ in Figs. \ref{fig:sample_1} and \ref{fig:sample_2} is remarkable and appears to represent the inherent limit of accuracy within the reference solution used in the calculation. The overall accuracy of \texttt{CNUCTRAN} improves as the final time step size decreases. It is important to note that only nuclides with concentrations above $10^{-30}$ were considered to minimize rounding errors. Additionally, the relative error increases with decreasing nuclide concentrations. Furthermore, almost all important nuclides have a relative error of less than $10^{-5} (0.001\%)$, indicating the outstanding accuracy of \texttt{CNUCTRAN}. For the benefit of the readers, we also provide CPU times for \texttt{CNUCTRAN} and the \texttt{CRAM48} method in Tables \ref{tbl:cpu1} and \ref{tbl:cpu2}. It is important to note that \texttt{CRAM48} outperforms \texttt{CNUCTRAN} in terms of the computational time, because it pre-computes the coefficients of the Chebychev approximants, $a$ and $\theta$, in Eq. (\ref{cram}). Meanwhile, \texttt{CNUCTRAN} prioritizes calculation accuracy over speed. However, its computational time is reasonable given modern computer speeds.

\begin{table}[h]
	\centering
	\small
	\caption{CPU time (seconds) comparison between two codes for the first sample calculation, performed with a 2.6GHz CPU.}
	\small
	\begin{tabular}{lllll}
		\hline
		final time step, $t_f$ & \texttt{CNUCTRAN} CPU Time &  \texttt{CRAM48} CPU Time &  &  \\
		\hline
		30 days                                                     & 1.9128                                                                                & 0.5938                                                                            &  &  \\
		One year                                                      & 2.1052                                                                                & 0.6084                                                                            &  &  \\
		30 years                                                      & 2.2028                                                                                & 0.6250                                                                            &  &  \\
		1 million years                                               & 2.4686                                                                                & 0.7125                                                                            &  & \\
		\hline
	\end{tabular}
	\label{tbl:cpu1}
\end{table}
\begin{table}[h]
	\centering
	\caption{CPU time (s) comparison between two codes for the second sample calculation, performed with a 2.6GHz CPU.}
	\begin{tabular}{lllll}
		\hline
		final time step, $T$    & \texttt{CNUCTRAN} CPU Time & \texttt{CRAM48} CPU Time&  &  \\
		\hline
		30 days       & 4.4719                                                                                & 0.7656                                                                            &  &  \\
		One year        & 4.9802                                                                                & 0.8281                                                                            &  &  \\
		30 years        & 5.1110                                                                                & 0.9844                                                                            &  &  \\
		1 million years & 5.9203                                                                                & 1.1406                                                                            &  & \\
		\hline
	\end{tabular}
   \label{tbl:cpu2}
\end{table}

\section{Conclusion}
In summary, this paper introduces the \texttt{CNUCTRAN} program, implementing a novel approach to solving nuclide burnup problems using a probabilistic method. The new method offers a unique alternative to traditional matrix exponential calculations and provides high computational fidelity. By adopting a probabilistic approach, \texttt{CNUCTRAN} successfully tackles the challenges of solving complex nuclide burnup problem without actually solving the associated Bateman equations. The code has been verified for two sample scenarios: simple decay and combined reactor and decay problems. The results demonstrated the accuracy and efficiency of \texttt{CNUCTRAN}. In the first sample, \texttt{CNUCTRAN} showed a high level of consistency with the Chebyshev Rational Approximation method (CRAM). \texttt{CNUCTRAN}'s results were compared with CRAM of order 48 (\texttt{CRAM48}) for all sample calculations, showing negligible differences and remarkable accuracy. Here, the relative errors between \texttt{CNUCTRAN} and CRAM are mostly $<10^{-5} (0.001\%)$ for all important nuclides with a fractional concentration of $\ge10^{-15}$. CRAM performs better than \texttt{CNUCTRAN} in terms of computation time, however, the average CPU time of \texttt{CNUCTRAN} is reasonably practical, allowing the focus on the computational fidelity rather than the speed. \texttt{CNUCTRAN} has proven its readiness for burnup calculations across various dimensional transmutation analyses. It offers a user-friendly input structure in XML format and provides the flexibility to adjust various simulation parameters. \hlr{In the future, \texttt{CNUCTRAN}} could be improved to allow users to specify neutron flux or power density. This will further enhance its applicability to practical situations.

\section{Declaration of competing interest}
\noindent The authors declare that they have no known competing financial interests or personal relationships that could have appeared to influence the work reported in this paper. 
* Items marked with an asterisk are only required for new versions
of programs previously published in the CPC Program Library.\\
\end{small}

\section{Acknowledgment}
\noindent This work was supported by the Ministry of Higher Education Malaysia through the Fundamental Research Grant Scheme [Project Code: FRGS/1/2022/STG07/USM/02/9]. J. Y. H. Soo acknowledges financial support via the Fundamental Research Grant Scheme (FRGS) by the Malaysian Ministry of Higher Education [Project code: FRGS/1/2023/STG07/USM/02/14].





\bibliographystyle{elsarticle-num}
\bibliography{refs.bib}

\begin{thebibliography}{10}
\expandafter\ifx\csname url\endcsname\relax
  \def\url#1{\texttt{#1}}\fi
\expandafter\ifx\csname urlprefix\endcsname\relax\def\urlprefix{URL }\fi
\expandafter\ifx\csname href\endcsname\relax
  \def\href#1#2{#2} \def\path#1{#1}\fi

\bibitem{Omar2022}
M.~R. Omar, J.~A. Karim, {The probability distribution of transmutation events
  and its application for solving burnup problems}, Progress in Nuclear Energy
  152 (Oct. 2022).
\newblock \href {https://doi.org/10.1016/j.pnucene.2022.104395}
  {\path{doi:10.1016/j.pnucene.2022.104395}}.

\bibitem{Origen1980}
A.~G. Croff, User's manual for the origen2 computer code, Tech. rep., Oak Ridge
  National Lab. (1980).

\bibitem{Monteburns}
H.~R. Trellue, \href{https://www.osti.gov/biblio/2696}{Development of
  monteburns: A code that links mcnp and origen2 in an automated fashion for
  burnup calculations}, Tech. rep. (12 1998).
\newblock \href {https://doi.org/10.2172/2696} {\path{doi:10.2172/2696}}.
\newline\urlprefix\url{https://www.osti.gov/biblio/2696}

\bibitem{Serpent2015}
J.~Leppänen, M.~Pusa, T.~Viitanen, V.~Valtavirta, T.~Kaltiaisenaho,
  \href{https://www.sciencedirect.com/science/article/pii/S0306454914004095}{The
  serpent monte carlo code: Status, development and applications in 2013},
  Annals of Nuclear Energy 82 (2015) 142--150, joint International Conference
  on Supercomputing in Nuclear Applications and Monte Carlo 2013, SNA + MC
  2013. Pluri- and Trans-disciplinarity, Towards New Modeling and Numerical
  Simulation Paradigms.
\newblock \href {https://doi.org/https://doi.org/10.1016/j.anucene.2014.08.024}
  {\path{doi:https://doi.org/10.1016/j.anucene.2014.08.024}}.
\newline\urlprefix\url{https://www.sciencedirect.com/science/article/pii/S0306454914004095}

\bibitem{Cinder1995}
T.~R. England, Cinder--a one point depletion and fission product program, Tech.
  rep., Westinghouse Electric Corp. Bettis Atomic Power Lab., West Mufflin,
  Penna. (1962).

\bibitem{Bowman1995}
S.~M. Bowman, M.~D. Dehart, C.~V. Parks, {Validation of SCALE-4 for Burnup
  Credit Applications}, Nucl Technol 110~(1) (1995) 53–70.
\newblock \href {https://doi.org/10.13182/NT95} {\path{doi:10.13182/NT95}}.

\bibitem{Aleph2007}
W.~Haeck, B.~Verboomen, \href{https://doi.org/10.13182/NSE07-A2695}{An optimum
  approach to monte carlo burnup}, Nuclear Science and Engineering 156~(2)
  (2007) 180--196.
\newblock \href {http://arxiv.org/abs/https://doi.org/10.13182/NSE07-A2695}
  {\path{arXiv:https://doi.org/10.13182/NSE07-A2695}}, \href
  {https://doi.org/10.13182/NSE07-A2695} {\path{doi:10.13182/NSE07-A2695}}.
\newline\urlprefix\url{https://doi.org/10.13182/NSE07-A2695}

\bibitem{Zhao2021}
Z.~Zhao, Y.~Yang, Q.~Gao, {Development and validation of Burn-up Calculation
  Code IMPC-Burnup2.0 for accelerator-driven sub-critical system}, Comput Phys
  Commun 261 (Apr. 2021).
\newblock \href {https://doi.org/10.1016/j.cpc.2020.107343}
  {\path{doi:10.1016/j.cpc.2020.107343}}.

\bibitem{Cetnar2021}
J.~Cetnar, P.~Stanisz, M.~Oettingen, Linear chain method for numerical
  modelling of burnup systems (2021).
\newblock \href {https://doi.org/10.3390/en14061520}
  {\path{doi:10.3390/en14061520}}.

\bibitem{Pusa2013}
M.~Pusa, J.~Leppänen, \href{https://doi.org/10.13182/NSE12-52}{Solving linear
  systems with sparse gaussian elimination in the chebyshev rational
  approximation method}, Nuclear Science and Engineering 175 (2013) 250--258,
  doi: 10.13182/NSE12-52.
\newblock \href {https://doi.org/10.13182/NSE12-52}
  {\path{doi:10.13182/NSE12-52}}.
\newline\urlprefix\url{https://doi.org/10.13182/NSE12-52}

\bibitem{Tavakkoli2021}
E.~Tavakkoli, M.~Zangian, A.~Minuchehr, A.~Zolfaghari, {Development and
  validation of ISOBURN, a new depletion code}, Ann Nucl Energy 159 (2021)
  108319.
\newblock \href {https://doi.org/10.1016/j.anucene.2021.108319}
  {\path{doi:10.1016/j.anucene.2021.108319}}.

\bibitem{Lanversin2021}
J.~de~Troullioud~de Lanversin, M.~Kütt, A.~Glaser, {ONIX: An open-source
  depletion code}, Ann Nucl Energy 151 (Feb. 2021).
\newblock \href {https://doi.org/10.1016/j.anucene.2020.107903}
  {\path{doi:10.1016/j.anucene.2020.107903}}.

\bibitem{Okamura2022}
T.~Okamura, R.~Katano, A.~Oizumi, K.~Nishihara, M.~Nakase, A.~Hidekazu,
  K.~Takeshita, Cost-reduced depletion calculation including short half-life
  nuclides for nuclear fuel cycle simulation, Journal of Nuclear Science and
  Technology 60~(6) (2023) 632--641.
\newblock \href {https://doi.org/10.1080/00223131.2022.2131646}
  {\path{doi:10.1080/00223131.2022.2131646}}.

\bibitem{Stankovskiy2012}
A.~Stankovskiy, G.~V. den Eynde,
  \href{https://doi.org/10.1155/2012/545103}{Advanced method for calculations
  of core burn-up, activation of structural materials, and spallation products
  accumulation in accelerator-driven systems}, Science and Technology of
  Nuclear Installations 2012 (2012) 545103.
\newblock \href {https://doi.org/10.1155/2012/545103}
  {\path{doi:10.1155/2012/545103}}.
\newline\urlprefix\url{https://doi.org/10.1155/2012/545103}

\bibitem{sublet2017}
J.~C. Sublet, J.~W. Eastwood, J.~G. Morgan, M.~R. Gilbert, M.~Fleming,
  W.~Arter, {FISPACT-II: An Advanced Simulation System for Activation,
  Transmutation and Material Modelling}, Nuclear Data Sheets 139 (2017)
  77–137.
\newblock \href {https://doi.org/10.1016/j.nds.2017.01.002}
  {\path{doi:10.1016/j.nds.2017.01.002}}.

\bibitem{Omar2024}
M.~Omar, W.~{Wan Mokhtar},
  \href{https://www.sciencedirect.com/science/article/pii/S030645492300484X}{Direct
  simulation method for estimating the concentration of nuclides undergoing
  transmutations}, Annals of Nuclear Energy 195 (2024) 110165.
\newblock \href {https://doi.org/https://doi.org/10.1016/j.anucene.2023.110165}
  {\path{doi:https://doi.org/10.1016/j.anucene.2023.110165}}.
\newline\urlprefix\url{https://www.sciencedirect.com/science/article/pii/S030645492300484X}

\bibitem{Muller2018}
E.~Müller,
  \href{https://www.sciencedirect.com/science/article/pii/S0306454918303141}{Essentially
  nonnegative matrix exponential methods for nuclide transmutation}, Annals of
  Nuclear Energy 120 (2018) 611--624.
\newblock \href {https://doi.org/https://doi.org/10.1016/j.anucene.2018.06.012}
  {\path{doi:https://doi.org/10.1016/j.anucene.2018.06.012}}.
\newline\urlprefix\url{https://www.sciencedirect.com/science/article/pii/S0306454918303141}

\bibitem{Hykes2013}
J.~M. Hykes, R.~M. Ferrer, Solving the bateman equations in casmo5 using
  implicit ode numerical methods for stiff systems, Proc. M\&C (2013).

\bibitem{Lahaye2017}
S.~Lahaye, A.~Tsilanizara, P.~Bellier, T.~Bittar, Implementation of a cram
  solver in mendel depletion code system, in: M\&C-2017 International
  Conference on Mathematics and Computational Methods Applied to Nuclear
  Science and Engineering, 2017.

\bibitem{lago2017}
D.~Lago, F.~Rahnema, {Development of a set of benchmark problems to verify
  numerical methods for solving burnup equations}, Ann Nucl Energy 99 (2017)
  266–271.
\newblock \href {https://doi.org/10.1016/j.anucene.2016.09.004}
  {\path{doi:10.1016/j.anucene.2016.09.004}}.

\bibitem{pusa2016}
M.~Pusa, {Higher-order Chebyshev rational approximation method and application
  to burnup equations}, Nuclear Science and Engineering 182~(3) (2016)
  297–318.
\newblock \href {https://doi.org/10.13182/NSE15-26}
  {\path{doi:10.13182/NSE15-26}}.

\bibitem{Gordon1998}
D.~M. Gordon,
  \href{https://www.sciencedirect.com/science/article/pii/S0196677497909135}{A
  survey of fast exponentiation methods}, Journal of Algorithms 27 (1998)
  129--146.
\newblock \href {https://doi.org/https://doi.org/10.1006/jagm.1997.0913}
  {\path{doi:https://doi.org/10.1006/jagm.1997.0913}}.
\newline\urlprefix\url{https://www.sciencedirect.com/science/article/pii/S0196677497909135}

\bibitem{Gustavson1978}
F.~G. Gustavson, \href{https://doi.org/10.1145/355791.355796}{Two fast
  algorithms for sparse matrices: Multiplication and permuted transposition},
  ACM Trans. Math. Softw. 4~(3) (1978) 250–269.
\newblock \href {https://doi.org/10.1145/355791.355796}
  {\path{doi:10.1145/355791.355796}}.
\newline\urlprefix\url{https://doi.org/10.1145/355791.355796}

\bibitem{Chadwick2011}
M.~Chadwick, M.~Herman, P.~Obložinský, M.~Dunn, Y.~Danon, A.~Kahler,
  D.~Smith, B.~Pritychenko, G.~Arbanas, R.~Arcilla, R.~Brewer, D.~Brown,
  R.~Capote, A.~Carlson, Y.~Cho, H.~Derrien, K.~Guber, G.~Hale, S.~Hoblit,
  S.~Holloway, T.~Johnson, T.~Kawano, B.~Kiedrowski, H.~Kim, S.~Kunieda,
  N.~Larson, L.~Leal, J.~Lestone, R.~Little, E.~McCutchan, R.~MacFarlane,
  M.~MacInnes, C.~Mattoon, R.~McKnight, S.~Mughabghab, G.~Nobre, G.~Palmiotti,
  A.~Palumbo, M.~Pigni, V.~Pronyaev, R.~Sayer, A.~Sonzogni, N.~Summers,
  P.~Talou, I.~Thompson, A.~Trkov, R.~Vogt, S.~van~der Marck, A.~Wallner,
  M.~White, D.~Wiarda, P.~Young, Endf/b-vii.1 nuclear data for science and
  technology: Cross sections, covariances, fission product yields and decay
  data, Nuclear Data Sheets 112 (2011) 2887--2996.
\newblock \href {https://doi.org/10.1016/J.NDS.2011.11.002}
  {\path{doi:10.1016/J.NDS.2011.11.002}}.

\end{thebibliography}


\end{document}